\documentclass[aps,prb,preprint,superscriptaddress]{revtex4-1}

\usepackage{amsmath}
\usepackage{graphicx}
\usepackage{placeins}

\mathchardef\mhyphen="2D
\newcommand{\beginsupplement}{ %
		\setcounter{figure}{0}
		\renewcommand{\thefigure}{S\arabic{figure}}%
		\setcounter{equation}{0}
		\renewcommand{\theequation}{S\arabic{equation}} %
		\setcounter{section}{0}
		\renewcommand{\thesection}{S\arabic{section}}
	}

\begin{document}

\title{Exciton Trapping Is Responsible for \\ the Long Apparent Lifetime in Acid-Treated MoS$_2$}

\author{Goodman, A.J.}
\affiliation{Department of Chemistry, Massachusetts Institute of Technology}
\author{Willard, A.P.}
\affiliation{Department of Chemistry, Massachusetts Institute of Technology}
\author{Tisdale, W.A.}
\affiliation{Department of Chemical Engineering, Massachusetts Institute of Technology}
\email[]{tisdale@mit.edu}

\date{\today}

\begin{abstract}
Here, we show that deep trapped ``dark'' exciton states are responsible for the surprisingly long lifetime of band-edge photoluminescence in acid-treated single-layer MoS$_2$. Temperature-dependent transient photoluminescence spectroscopy reveals an exponential tail of long-lived states extending hundreds of meV into the band gap. These sub-band states, which are characterized by a 4 $\mu$s radiative lifetime, quickly capture and store photogenerated excitons before subsequent thermalization up to the band edge where fast radiative recombination occurs. By intentionally saturating these trap states, we are able to measure the ``true'' 150 ps radiative lifetime of the band-edge exciton at 77~K, which extrapolates to $\sim$600 ps at room temperature. These experiments reveal the dominant role of dark exciton states in acid-treated MoS$_2$, and suggest that excitons spend $>95\%$ of their lifetime at room temperature in trap states below the band edge. We hypothesize that these states are associated with native structural defects, which are not introduced by the superacid treatment; rather, the superacid treatment dramatically reduces non-radiative recombination through these states, extending the exciton lifetime and increasing the likelihood of eventual radiative recombination. 
\end{abstract}

\maketitle

MoS$_{2}$ is a two-dimensional semiconductor that exhibits high carrier mobilities\cite{MoS2_transistor} and tightly-bound luminescent excitons centered at the direct gap at the K point in the Brilluoin zone in the monolayer limit.\cite{KFMAk_PRL2010,WS2_nonhydrogenic,RydbergStates_MoS2_Heinz_Nanolett2015} These properties, together with the ability to control spin-valley polarization,\cite{Valleytronics_XuReview_NatureReviews_2016} make MoS$_{2}$ a promising material for semiconducting devices in compact device architectures.\cite{MoS2_phototransistor_ACSNano2012,WSe2OptoElec,WSe2LED,MoS2_ExcitonicLED_APL2014,MoS2_Electroluminescence_NanoLett2013,MoS2_transistor_1nmGate} 

In a key discovery that will enable more efficient optoelectronic devices, Amani \textit{et al.} reported a chemical superacid treatment\cite{NearUnityPL_QY_MoS2_Amani_etal_Science2015} that increased the photoluminescence quantum yield (QY) of MoS$_2$ from $< 1\%$  as-exfoliated to $> 95 \%$. Simultaneously, they observed an increase in the photoluminescence lifetime from roughly 250 ps to 10 ns, which is itself surprising. As pointed out by Schaibley \textit{et al.}, the strong transition dipole moment in MoS$_2$ should lead to a short intrinsic radiative lifetime.\cite{Valleytronics_XuReview_NatureReviews_2016} Calculations predict a room temperature lifetime in the range of hundreds of picoseconds,\cite{excitonPolaritons_heterostructures_microcav,TMDC_ExcitonLifetime_Grossman_2015Nanolett} in agreement with some experiments.\cite{MoS2_2DElectronicSpectroscopy,UltrfastMoS2_TRPL_LowTemp_APL2011} However, many processes can obscure the actual radiative lifetime, such as defect-mediated non-radiative decay,\cite{FastAnnihilationDefects_TMDS_PRB2015} exciton-exciton interactions,\cite{UltrafastMoS2_XXAnnihilation_NL2014} or equilibration with dark states.\cite{DarkExcitonsWSe2_HeinzPRL2015} Here we show that the long apparent lifetime in acid-treated MoS$_2$ is due to the presence of long-lived (4 $\mu$s) dark states that extend hundreds of meV into the bandgap. At room temperature, excitons spend the majority of their lifetimes in these dark states before thermalizing to the band edge and emitting with the free exciton radiative rate. We measure the true radiative lifetime at 77~K to be 150~ps by saturating the traps and observing free exciton emission.

\section{Results and Discussion}
\begin{figure}[h]
\includegraphics[width=.45\columnwidth]{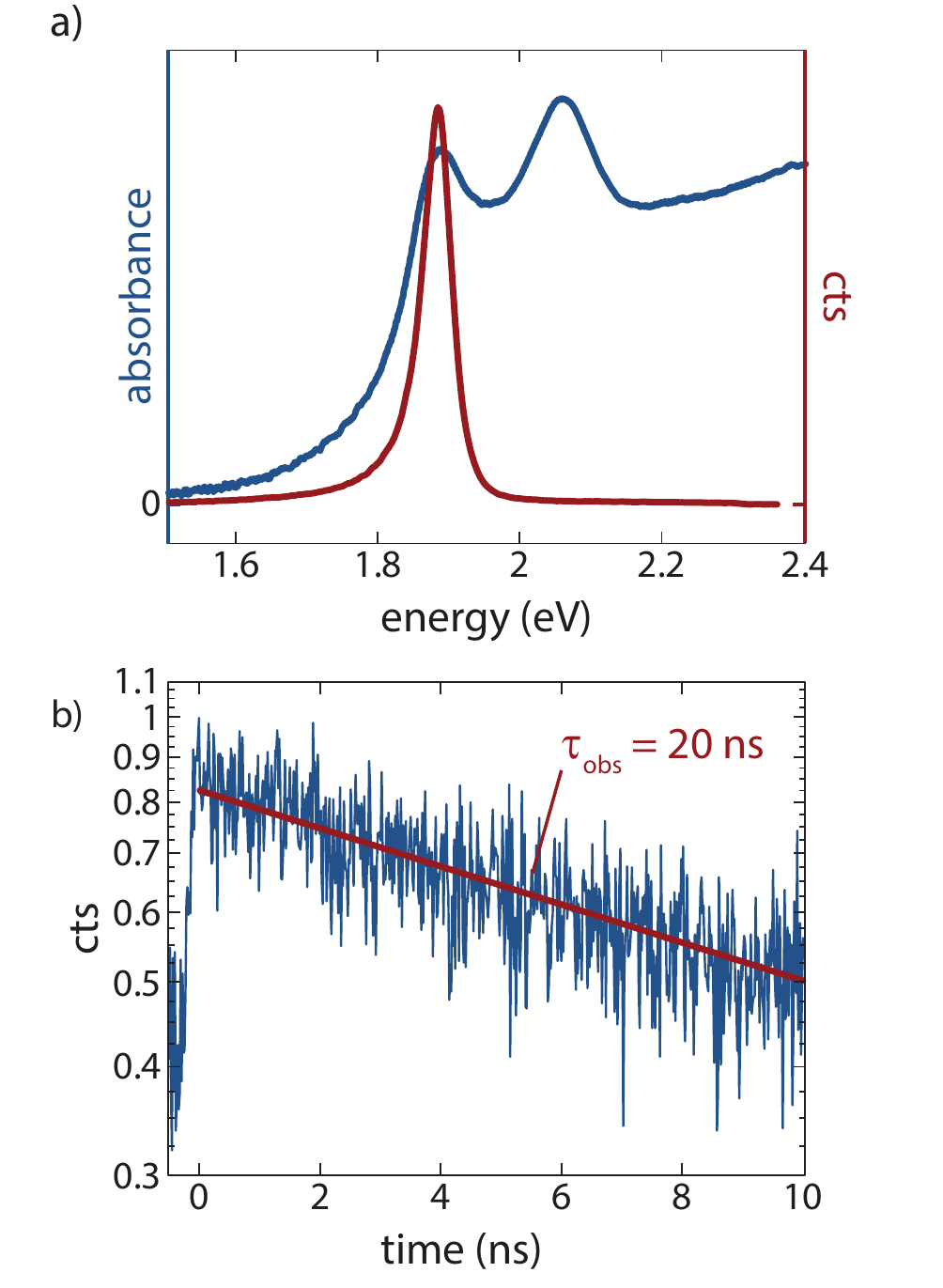}
\caption{Acid-treated MoS$_2$ optical properties at room temperature. (a) The absorbance spectrum (blue) shows two sharp, excitonic resonances arising from the A and B excitons that originate from the K point in the Brilluoin zone. (b) Time-correlated single photon counting histogram of the A exciton photoluminescence at room temperature exhibits single-exponential decay dynamics with a time constant, $\tau_\mathrm{obs}=20~\mathrm{ns}$.}
\label{figure1}
\end{figure}

Absorption and photoluminescence spectroscopy reveal the high material quality achievable with the \textit{bis}(trifluoromethane)sulfonimide (TFSI) superacid treatment described by Amani \textit{et al.} The absorbance and emission spectra of acid-treated MoS$_2$ are plotted in Fig.~\ref{figure1}. The absorbance spectrum in Fig.~\ref{figure1}a exhibits two narrow absorbance features corresponding to the A and B excitons originating from the spin-split conduction band at the K point of the Brilluoin zone. The peaked absorbance reflects the excitonic nature of the transition. The red line shows the narrow photoluminescence spectrum exhibiting clean luminescence from the A exciton.  Fig.~\ref{figure1}b shows the photoluminescence decay dynamics. The data were collected at sufficiently low fluence to avoid exciton-exciton interactions as evidenced by the monoexponential decay. The observed lifetime is roughly 20~ns, consistent with prior work.

One can estimate what radiative lifetimes in 2D quantum wells should be using a relatively simple model\cite{IntrinsicRadDecayTimes_quantumWells} that translates well to TMDs.\cite{IntrinsicRadDecayTimes_2DMaterials,2D_Mat_RadLifetimes_PRB} Following the approach presented by Robert \textit{et al.}, we can estimate the intrinsic radiative lifetime in MoS$_2$ using the equation
\begin{align}
\tau_{\mathrm{rad}}^0=\frac{\hbar\epsilon}{2k_0}\left(\frac{E_{X^0}}{e\hbar v}\right)^2\left(a_B\right)^2,
\label{tau_intrins}
\end{align}
where $\epsilon$ is the dielectric constant of MoS$_2$, $k_0$ is the magnitude of the wavevector of exciton emission, $E_{X^0}$ is the exciton energy, $v$ the Kane velocity, and $a_B$ the exciton Bohr radius. However, at finite temperature, only a small subset of a thermalized exciton population can emit while satisfying momentum conservation. This effect modifies the radiative rate according to
\begin{align}
\tau_{\mathrm{rad}}^{\mathrm{eff}}=\frac{3}{2}\frac{k_BT}{E}\tau_{\mathrm{rad}}^0,
\label{taueff}
\end{align}
where $E$ is a bound on the kinetic energy an exciton can possess and still emit radiatively. Using values for the MoS$_2$ dielectric constant\cite{TMD_dielectric_measured_HeinzPRB2014} and exciton Bohr radius\cite{CalcBindingEnergy_X_XX_MoS2_PRB2015} from the literature, equation~(\ref{taueff}) implies a room temperature exciton lifetime of 470 ps, which is more than forty times smaller than the measured lifetime evident in Fig.~\ref{figure1}b. This is in reasonable agreement with the even shorter 270 ps room temperature radiative lifetime computed with first principles calculations.\cite{TMDC_ExcitonLifetime_Grossman_2015Nanolett} 

\begin{figure}[h]
\includegraphics[width=.5\columnwidth]{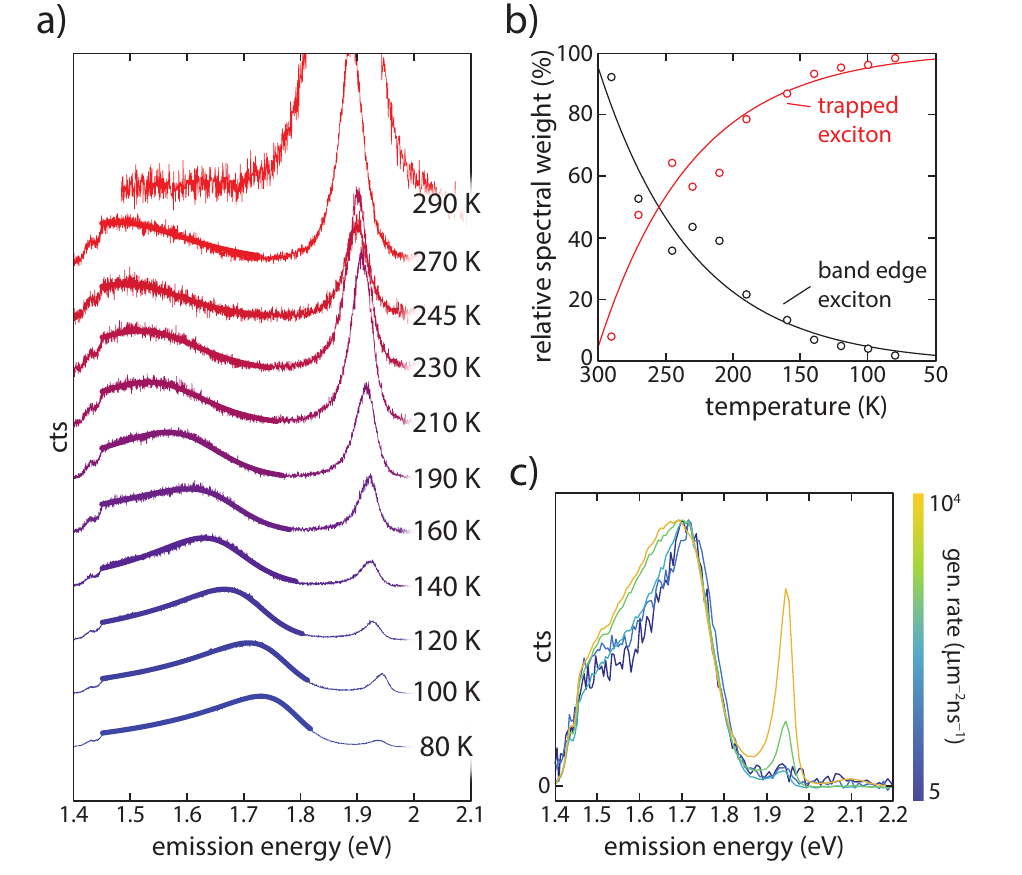}
\caption{Temperature-dependent photoluminescence spectra (a) Photoluminescence spectra (thin lines) at low temperature exhibit emission at energies below the band-edge exciton. The lineshape is fit well by an exponential density of states occupied by a Fermi-Dirac Distribution (thick lines). (b) At low temperature, trapped exciton emission dominates the photoluminescence spectra, while at high temperatures most luminescence comes from the band-edge exciton. (c) Power-dependent photoluminescence spectra at 77 K reveal that the band-edge exciton emission becomes more prominent relative to the trapped exciton emission at higher laser powers. This observation indicates that a significant portion of the traps are filled at moderate excitation intensities.}
\label{figure2}
\end{figure}

We believe long-lived trapped excitons explain the difference between the expected and measured decay times. Crystallographic defects are known to permeate exfoliated MoS$_2$,\cite{IntrinsicStructural_MonolayerMoS2_NanoLett,MoS2_SurfaceDefectsOnNatural_ACSAMI_2015} and persist even in well passivated acid-treated MoS$_2$.\cite{NearUnityPL_QY_MoS2_Amani_etal_Science2015,Amani_TFSI_Nanoletters2016_followup} Though not evident in room temperature absorbance or photoluminescence spectra, long-lived emission from sub-band traps comprises a large fraction of the photoluminescence spectrum at cryogenic temperatures. Fig.~\ref{figure2}a presents photoluminescence spectra acquired at many temperatures between 77 K and 300 K. The spectra are normalized to the maximum emission intensity below the band-edge exciton to highlight and compare trapped exciton emission. The percentage of the total emission intensity from the trapped excitons is presented in Fig.~\ref{figure2}b. At high temperature, nearly all of the emission comes from the band-edge exciton, whereas at 77 K, nearly all emission comes from trapped excitons. The intensity and spectral distribution of trap state emission varies from sample to sample, but its presence is observed in both exfoliated MoS$_2$ and CVD-grown samples (see Supporting Information).

The trapped exciton emission lineshape is well represented by a Fermi-Dirac distribution over an exponential tail of states extending into the band gap (thick lines), as shown in Fig.~\ref{figure2}a. The density of traps states, $\rho\propto\exp\left[-\alpha\left(E-E_\textrm{band~endge}\right)\right]$, is parametrized by $\alpha\sim 5~\mathrm{eV}^{-1}$. We note that in all cases the effective temperature of the trapped exciton distribution is higher than the lattice temperature, indicating that the exciton population and lattice do not reach thermal equilibrium over the course of the exciton lifetime. This is rationalized by a model in which trap states -- each having a well-defined energy -- are spatially separated from one another. In this case, complete thermalization of the trapped exciton population is only achieved if the exciton is able to sample many different trap sites during its lifetime through multiple trapping/de-trapping and transport events. We motivate the use of a Fermi-Dirac distribution by noting that the trapped-exciton emission saturates at moderate laser power. Fig.~\ref{figure2}c shows the 77 K emission excited under laser power spanning 4 orders of magnitude. At higher steady-state exciton population densities, fewer trap sites are available for exciton trapping, making band edge emission more likely.

\begin{figure}[h!]
\includegraphics[width=.45\columnwidth]{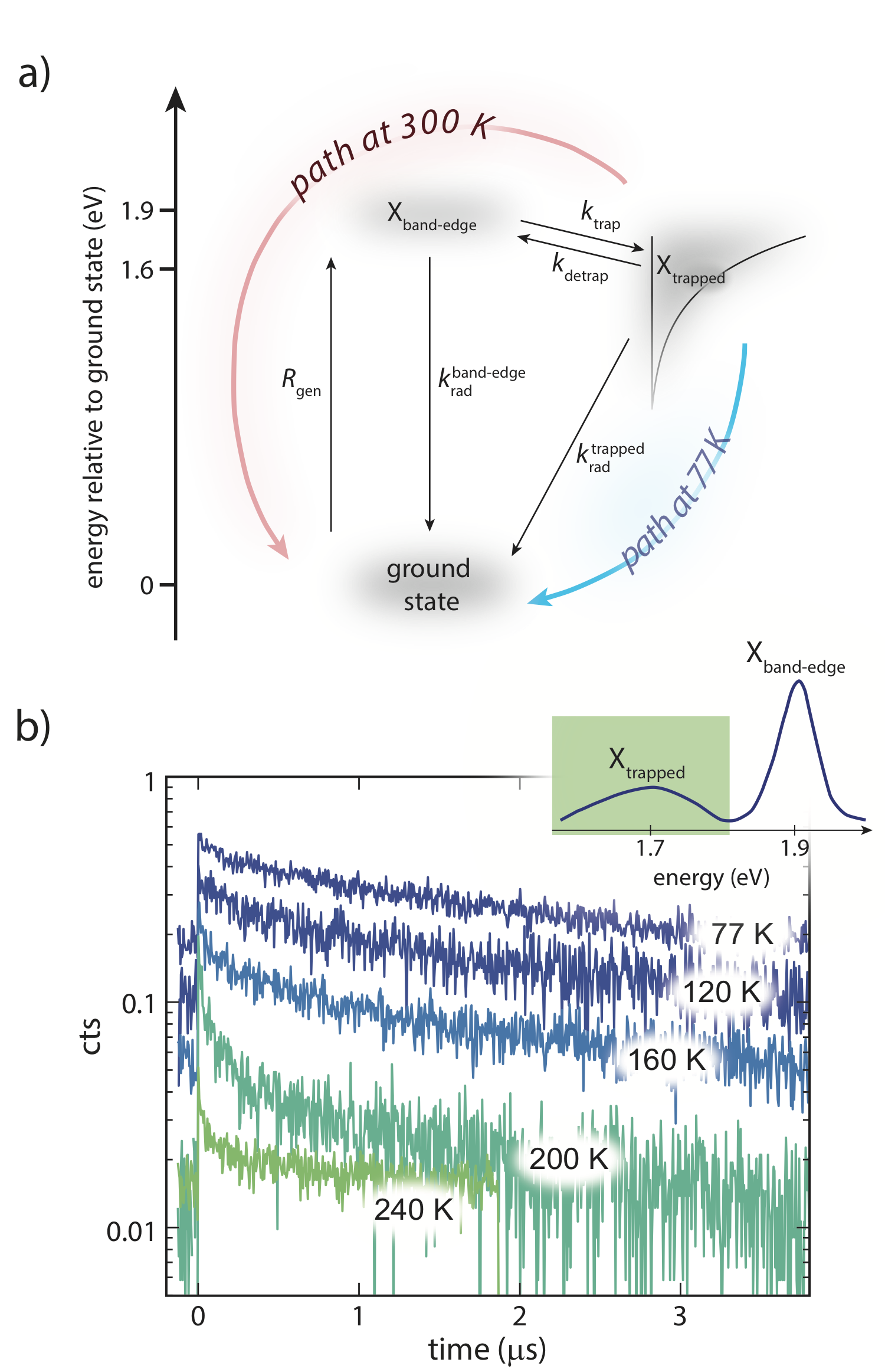}
\caption{A three-state model explains the observed photoluminescence decay time and the absence of trap state emission at room temperature. (a) Apart from the ground state and the band-edge exciton, there are also trapped exciton states that have $\sim$1000$\times$ slower recombination rate than the band-edge exciton. At room temperature, thermalization up to the band edge and subsequent radiative recombination is most probable, whereas at 77~K direct recombination from trap to ground state is observed. (b) Time-resolved trapped exciton emission reveals these two pathways. At 77 K, trapped excitons decay with their slow radiative rate $\left(\sim 4~\mathrm{\mu s}\right)$. At higher temperatures, some thermalize up to the band-edge leaving behind only deeply trapped excitons, which decay slowly. The shaded region of the inset indicates the portion of the spectrum that is collected by the time-resolved detector.}
\label{figure3}
\end{figure}

Fig.~\ref{figure3}a illustrates the paths that a trapped exciton takes in this energy landscape. Following photoexcitation, band edge excitons are quickly captured into trap states. From there, trapped excitons can either decay radiatively to the ground state directly (the blue path in Fig.~\ref{figure3}a) or use some of the available thermal energy to reach the band-edge before quickly decaying with the fast band-edge radiative rate (the red path in Fig.~\ref{figure3}a). We probe these processes by monitoring the emission dynamics from the trap state (using a long-pass filter to isolate the trap state emission) at a set of temperatures between 77~K and 240~K. The results of these measurements are presented in Fig.~\ref{figure3}b. At 77 K, there is not sufficient thermal energy to promote trapped excitons to the band edge; consequently, excitons decay radiatively from their trapped state with the trapped exciton lifetime ($\sim 4 ~\mathrm{\mu s}$). In contrast, at 240~K there is sufficient thermal energy present to promote a portion of the trapped exciton population to the band-edge with rate $k_{\mathrm{detrap}}=k_{\mathrm{trap}}\exp\left(-\Delta E/k_B T\right)$. This fast detrapping appears as fast decay components in the trap-state emission dynamics at high temperatures. The slow decay component represents trapped excitons that are too deeply trapped to be thermally activated to the band-edge. Intermediate temperatures exhibit a spectrum of detrapping rates reflecting the distribution of trapped exciton energies.

Using the three-state model illustrated in Fig.~\ref{figure3}a, we can construct the rate equations governing band-edge and trapped exciton dynamics. Those equations are:
\begin{widetext}
\begin{align}
\frac{\mathrm{d}[X]}{\mathrm{d}t}&= R_\mathrm{gen}-k_{\mathrm{rad}}^{\mathrm{band\mhyphen edge}}[X]
-k_{\mathrm{trap}}[X]\left(1-\frac{[X_T]}{N_0}\right)+k_{\mathrm{detrap}}[X_T]
\label{X_pop}\\
\frac{\mathrm{d}[X_T]}{\mathrm{d}t}&=-k_{\mathrm{rad}}^{\mathrm{trapped}}[X_T]
-k_{\mathrm{detrap}}[X_T]+k_{\mathrm{trap}}[X]\left(1-\frac{[X_T]}{N_0}\right)
\label{XT_pop}
\end{align}
\end{widetext}
where $[X]$ and $[X_T]$ are the band-edge and trapped exciton densities respectively, $R_\mathrm{gen}$, $k_{\mathrm{rad}}^{\mathrm{band\mhyphen edge}}$, $k_{\mathrm{rad}}^{\mathrm{trapped}}$, $k_{\mathrm{trap}},~\&~k_{\mathrm{detrap}}$ are the rates of generation, band-edge radiative decay, trapped exciton decay, trapping and detrapping respectively, and $N_0$ is the average area density of traps. The model represented by equations~(\ref{X_pop}-\ref{XT_pop})  captures many salient optical properties of MoS$_2$. The absence of non-radiative decay pathways reflects the near-unity QY reported by Amani \textit{et al.}\cite{NearUnityPL_QY_MoS2_Amani_etal_Science2015} The term $\left(1-[X_T]/N_0\right)$ represents the fraction of trap sites that are currently unoccupied and reproduces saturation behavior as the generation rate is increased. The trapping rate and  trap density can be found quantitatively by examining the data under steady-state continuous excitation (Fig.~\ref{figure2}c) in the context of equations~(\ref{X_pop}-\ref{XT_pop}) at equilibrium when their left-hand-sides are zero. Assuming thermally-activated detrapping and using the $\left(k_{\mathrm{rad}}^{\mathrm{trapped}}\right)^{-1}=4~\mu s$ measured at 77 K (see Fig.~\ref{figure3}b), we find $k_{\mathrm{trap}}=3\times10^{11}~\mathrm{s}^{-1}$ and $N_0=4\times10^6~\mathrm{\mu m}^{-2}$. With the same parameters we can fit the out-of-equilibrium, transient data presented in Fig.~\ref{figure3}b using a numerical ODE solver (see Supplementary Information).

\begin{figure}[h]
\includegraphics[width=.45\columnwidth]{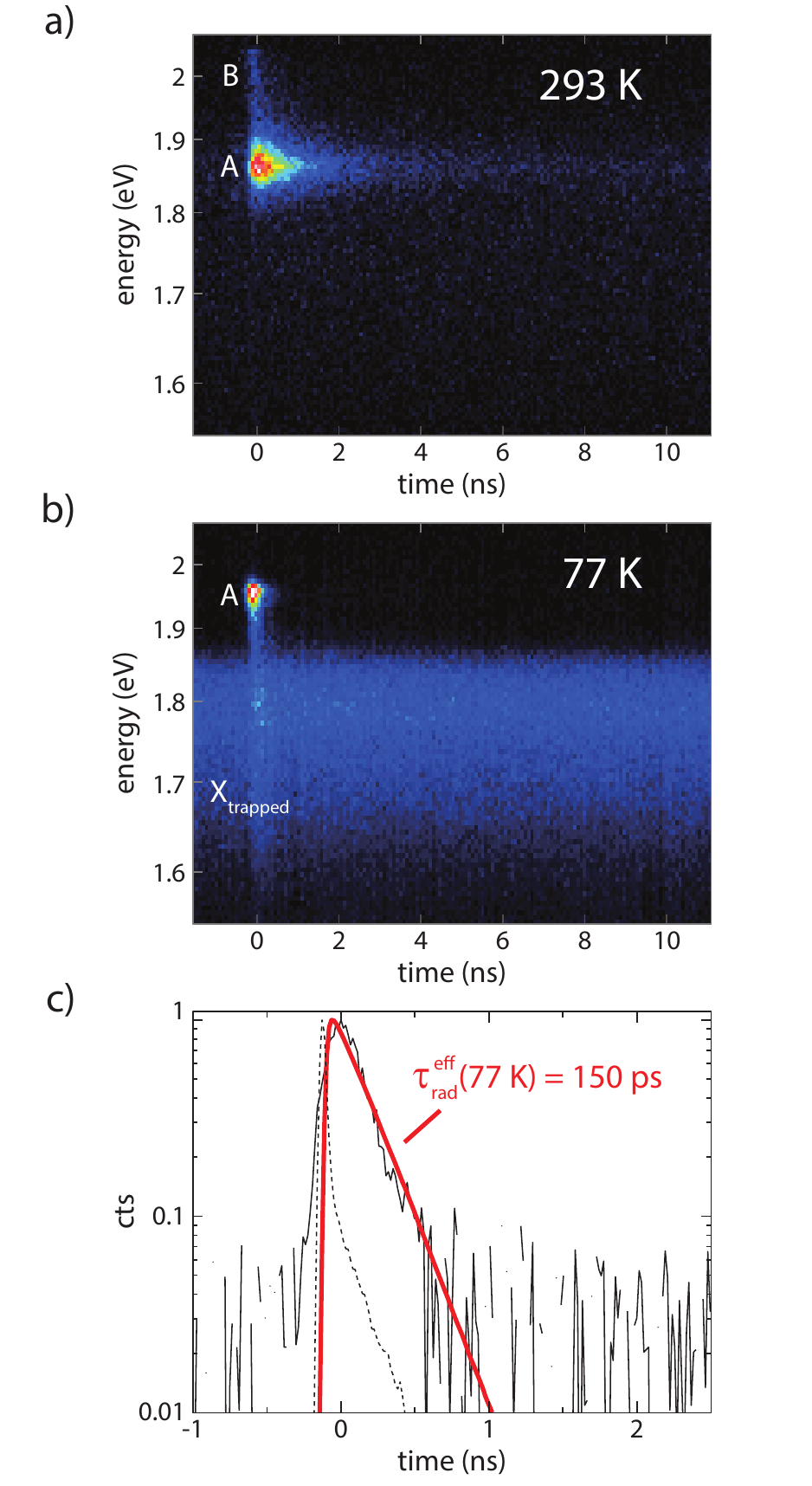}
\caption{Spectrally-resolved transient photoluminescence spectroscopy reveals the true radiative rate. (a) At room temperature, nearly all of the emission comes from the band-edge exciton, with a noticeable contribution from the B exciton. The exciton emission has long-lived components mediated by the long-lived, dark trapped exciton states. (b, c) At 77~K, the majority of emission comes from trapped excitons ($\sim$$1.65-1.85$ eV). The pulsed laser operates at 76 MHz (13~ns between pulses), so the long-lived trapped-exciton sites are always nearly completely occupied in this experiment. When a new pulse arrives, band edge excitons cannot trap and instead decay radiatively with a rate reflecting the true band-edge radiative rate, which is 150~ps at 77~K.}
\label{figure4}
\end{figure}

Spectrally-resolved transient photoluminescence spectroscopy reveals the final relevant rate constant, $k_{\mathrm{rad}}^{\mathrm{band\mhyphen edge}}$. Fig.~\ref{figure4}a shows the time-resolved photoluminescence spectrum recorded at room temperature. Nearly all of the emission is from the band-edge exciton with some initial contribution from the B exciton, consistent with steady-state photoluminescence spectra. 

Fig.~\ref{figure4}b shows the time-resolved photoluminescence spectrum collected at a sample temperature of 77~K. In addition to the blue-shifted A exciton emission, there is prominent, long-lived emission from the trapped exciton states. These data were collected with a 76 MHz pulsed laser. During the time between pulses, the trap state emission is nearly static with the exception of a small, fast component due to hot trapped-exciton luminescence. In this experiment, the trap sites are always nearly completely saturated and at equilibrium with the band-edge exciton state. When the next laser pulse arrives, no empty sites are available for trapping and the only available relaxation pathway is direct radiative recombination to the ground state. Consequently, the fast decay of the A exciton emission reflects the temperature-adjusted intrinsic radiative lifetime. Fig~\ref{figure4}c shows the decay dynamics of the A exciton emission integrated between 1.9 and 2 eV along with the experiment's impulse response function (IRF). The red line shows a single exponential convolved with the IRF and fitted to the experimental data. The extracted radiative lifetime is 150 ps. Extrapolating to room temperature using eqn.~(\ref{taueff}) this implies a room temperature lifetime $\tau_\mathrm{rad}^{\mathrm{eff}}(300~\mathrm{K})=580~\mathrm{ps}$ in reasonable agreement with the estimate made using eqns.~(\ref{tau_intrins})~$\&$~(\ref{taueff}). The observed 20 ns exciton lifetime at room temperature (Fig.~\ref{figure1}b) is $\sim$20 times longer than the lifetime of a free band-edge exciton, indicating that excitons in acid-treated MoS$_2$ spend $>95\%$ of their lifetime occupying trap states below the band edge. 

These results, in combination with recent work in the field,\cite{2D_ionicLiquid_XYZ_2017Nanolett,Amani_TFSI_Nanoletters2016_followup,CYTOPStability}  leads us to speculate with regards to the mechanism of photoluminescence enhancement. Monolayer MoS$_2$ is susceptible to sulfur vacancies.\cite{Amani_TFSI_Nanoletters2016_followup,IntrinsicStructural_MonolayerMoS2_NanoLett} Consequently, as-prepared MoS$_2$ is often found to be n-type, with dopant electrons compensating the partial positive charge of the sulfur site vacancy. Atallah \textit{et al.} recently observed PL enhancement of MoS$_2$ similar to the effects of the superacid treatment by immersion in an ionic liquid --– but only if the sample was electrically connected to ground to allow release of the dopant electrons.\cite{2D_ionicLiquid_XYZ_2017Nanolett} This suggests that the action of the superacid is to simultaneously passivate the structural defect with the conjugate base anion (TFSI$^-$) and remove dopant electrons \textit{via} chemical reduction of the acidic proton $(2\textrm{H}^+ + 2\textrm{e}^- \rightarrow \textrm{H}_2)$. Such a hypothesis is consistent with Javey and coworkers' recent demonstration of enhanced acid-treated MoS$_2$ stability in the presence of a capping fluoropolymer (protecting the residual surface-bound TFSI anion), and the superacid's tendency to make n-doped materials more intrinsic.\cite{NearUnityPL_QY_MoS2_Amani_etal_Science2015}

In agreement with this speculation, we hypothesize that the trap states observed here are not introduced by the superacid treatment. Rather, we believe they are associated with native structural defects (most likely sulfur vacancies) that survive the acid treatment.\cite{Amani_TFSI_Nanoletters2016_followup} In as-exfoliated samples, these trap states act as recombination centers increasing non-radiative recombination and reducing the QY. The action of the acid treatment is to dramatically reduce the rate of non-radiative recombination from trap-to-ground state, which together with the intrinsically slow radiative rate of this transition, leads to longer exciton lifetimes. Consequently, the superacid treatment does nothing but improve the photophysical properties of MoS$_2$. To achieve trap-free exciton dynamics limited only by intrinsic radiative recombination, more pristine starting materials are needed.

This work demonstrates that trapped excitons significantly affect room temperature exciton dynamics even though they don't appear in absorbance or photoluminescence spectra at room temperature. The trapped excitons possess small transition dipole moments and degeneracy relative to the band-edge exciton states, which explains their absence in room temperature measurements. That these traps extend so far into the bandgap, however, ensures that these states play a role in the lives of excitons at room and elevated temperatures. It is important to understand the true energy landscape seen by excitons in acid-treated MoS$_2$ as the presence of deep traps could affect exciton transport and dynamics, processes critical to most excitonic optoelectronic devices. With better understanding of the identity, energy, and distribution of these traps, it may be possible to alter the trap state energy distribution using the external dielectric environment or adjust the trap occupancy \textit{via} electrostatic doping.

\FloatBarrier

\section{methods}
\subsection{Sample Preparation}
Flakes were prepared by mechanically exfoliating an MoS$_2$ single crystal (Graphene Supermarket). For most measurements, samples were exfoliated to Si/SiO$_2$ to aid in monolayer identification. Sample thicknesses were confirmed with Raman spectroscopy.\cite{MoS2Raman} For absorbance measurements, flakes were exfoliated to a No.~1 glass coverslip. Samples were superacid treated following the procedure reported by Amani \textit{et al.}\cite{NearUnityPL_QY_MoS2_Amani_etal_Science2015} \textit{bis}(trifluoromethane)sulfonimide (TFSI) was dissolved into anhydrous 1,2-dichloroethane in a nitrogen glovebox forming a 0.2 mg/mL solution. Samples were immersed in a freshly prepared solution for 10 minutes. After removing the vial containing the sample from the glovebox, the sample was removed from solution and blown dry under nitrogen. The sample was then placed in a 100$^\circ$C oven and heated in air for 5 minutes.
\subsection{Steady State Spectroscopy}
Spatially-resolved spectroscopy was performed in an inverted microscope. For steady-state spectra, a CW diode laser (Coherent, Sapphire SF, 532 nm) was focused at the sample (Nikon, CFI S Plan Fluor ELWD, 40$\times$, 0.6 NA). Fluorescence was passed by a long-pass dichroic reflector and imaged at the entrance slit of a 0.5 m focal length spectrograph and dispersed onto a cooled charge-coupled device (Princeton Instruments, Pixis). Emission spectra were transformed from wavelength (nm) to photon energy (eV) using the Jacobian transformation.\cite{Jacobian_Transform} 

Absorbance spectra were taken by illuminating a transparent substrate from above with a broad-band incoherent source (tungsten halogen lamp) and recording spectra transmitted through the substrate-only as well as the substrate/sample stack.
\subsection{Time-Correlated Single Photon Counting}
A pulsed excitation source (76 MHz source: Coherent OPO, PP automatic, 76 MHz, $<$ 1 ps; 250 kHz source: Picoquant, LDH-D-C405M, 0.4 ns pulse duration) excited the sample. Fluorescence was detected with a silicon avalanche photodiode (Micro Photon Devices, PDM50, 50 ps resolution at the detection wavelength), which sent a voltage pulse to a counting board (Picoquant, PicoHarp 300). Time-resolved spectra were acquired by first passing the fluorescence through a spectrograph and measuring decay histograms at each monochromated spectral slice.
\subsection{Numerical Modelling}
Steady-state spectra were modelled by taking the differential eqns.~(\ref{X_pop})$~\&~$(\ref{XT_pop}) at equilibrium resulting in a system of nonlinear equations that were solved numerically using fsolve in Matlab. 

Time-dependent dynamics were modelled by numerically integrating eqns.~(\ref{X_pop})$~\&~$(\ref{XT_pop}) using the ODE solver in Matlab.

\begin{acknowledgments}
We thank  Der-Hsien Lien, Geun Ho Ahn, and Ali Javey for helpful discussions. We thank Kevin Bogaert for CVD-grown MoS$_2$. This work was supported as part of the Center for Excitonics, an Energy Frontier Research Center funded by the U.S. Department of Energy, Office of Science, Basic Energy Sciences (BES) under award number: DE-SC0001088 (MIT). A.J.G. was partially supported by the U.S. National Science Foundation Graduate Research Fellowship Program under Grant No. 1122374.
\end{acknowledgments}

\bibliographystyle{apsrev4-1}
\bibliography{TrapStateLib}

\begin{thebibliography}{31}%
\makeatletter
\providecommand \@ifxundefined [1]{%
 \@ifx{#1\undefined}
}%
\providecommand \@ifnum [1]{%
 \ifnum #1\expandafter \@firstoftwo
 \else \expandafter \@secondoftwo
 \fi
}%
\providecommand \@ifx [1]{%
 \ifx #1\expandafter \@firstoftwo
 \else \expandafter \@secondoftwo
 \fi
}%
\providecommand \natexlab [1]{#1}%
\providecommand \enquote  [1]{``#1''}%
\providecommand \bibnamefont  [1]{#1}%
\providecommand \bibfnamefont [1]{#1}%
\providecommand \citenamefont [1]{#1}%
\providecommand \href@noop [0]{\@secondoftwo}%
\providecommand \href [0]{\begingroup \@sanitize@url \@href}%
\providecommand \@href[1]{\@@startlink{#1}\@@href}%
\providecommand \@@href[1]{\endgroup#1\@@endlink}%
\providecommand \@sanitize@url [0]{\catcode `\\12\catcode `\$12\catcode
  `\&12\catcode `\#12\catcode `\^12\catcode `\_12\catcode `\%12\relax}%
\providecommand \@@startlink[1]{}%
\providecommand \@@endlink[0]{}%
\providecommand \url  [0]{\begingroup\@sanitize@url \@url }%
\providecommand \@url [1]{\endgroup\@href {#1}{\urlprefix }}%
\providecommand \urlprefix  [0]{URL }%
\providecommand \Eprint [0]{\href }%
\providecommand \doibase [0]{http://dx.doi.org/}%
\providecommand \selectlanguage [0]{\@gobble}%
\providecommand \bibinfo  [0]{\@secondoftwo}%
\providecommand \bibfield  [0]{\@secondoftwo}%
\providecommand \translation [1]{[#1]}%
\providecommand \BibitemOpen [0]{}%
\providecommand \bibitemStop [0]{}%
\providecommand \bibitemNoStop [0]{.\EOS\space}%
\providecommand \EOS [0]{\spacefactor3000\relax}%
\providecommand \BibitemShut  [1]{\csname bibitem#1\endcsname}%
\let\auto@bib@innerbib\@empty
\bibitem [{\citenamefont {Radisavljevic}\ \emph {et~al.}(2011)\citenamefont
  {Radisavljevic}, \citenamefont {Radenovic}, \citenamefont {Brivio},
  \citenamefont {Giacometti},\ and\ \citenamefont {Kis}}]{MoS2_transistor}%
  \BibitemOpen
  \bibfield  {author} {\bibinfo {author} {\bibfnamefont {B.}~\bibnamefont
  {Radisavljevic}}, \bibinfo {author} {\bibfnamefont {A.}~\bibnamefont
  {Radenovic}}, \bibinfo {author} {\bibfnamefont {J.}~\bibnamefont {Brivio}},
  \bibinfo {author} {\bibfnamefont {V.}~\bibnamefont {Giacometti}}, \ and\
  \bibinfo {author} {\bibfnamefont {A.}~\bibnamefont {Kis}},\ }\href@noop {}
  {\bibfield  {journal} {\bibinfo  {journal} {Nat. Nanotechnol.}\ }\textbf
  {\bibinfo {volume} {6}},\ \bibinfo {pages} {147} (\bibinfo {year}
  {2011})}\BibitemShut {NoStop}%
\bibitem [{\citenamefont {Mak}\ \emph {et~al.}(2010)\citenamefont {Mak},
  \citenamefont {Lee}, \citenamefont {Hone}, \citenamefont {Shan},\ and\
  \citenamefont {Heinz}}]{KFMAk_PRL2010}%
  \BibitemOpen
  \bibfield  {author} {\bibinfo {author} {\bibfnamefont {K.}~\bibnamefont
  {Mak}}, \bibinfo {author} {\bibfnamefont {C.}~\bibnamefont {Lee}}, \bibinfo
  {author} {\bibfnamefont {J.}~\bibnamefont {Hone}}, \bibinfo {author}
  {\bibfnamefont {J.}~\bibnamefont {Shan}}, \ and\ \bibinfo {author}
  {\bibfnamefont {T.}~\bibnamefont {Heinz}},\ }\href@noop {} {\bibfield
  {journal} {\bibinfo  {journal} {Phys. Rev. Lett.}\ }\textbf {\bibinfo
  {volume} {105}},\ \bibinfo {pages} {136805} (\bibinfo {year}
  {2010})}\BibitemShut {NoStop}%
\bibitem [{\citenamefont {Chernikov}\ \emph {et~al.}(2014)\citenamefont
  {Chernikov}, \citenamefont {Berkelbach}, \citenamefont {Hill}, \citenamefont
  {Rigosi}, \citenamefont {Li}, \citenamefont {Aslan}, \citenamefont
  {Reichman}, \citenamefont {Hybertsen},\ and\ \citenamefont
  {Heinz}}]{WS2_nonhydrogenic}%
  \BibitemOpen
  \bibfield  {author} {\bibinfo {author} {\bibfnamefont {A.}~\bibnamefont
  {Chernikov}}, \bibinfo {author} {\bibfnamefont {T.}~\bibnamefont
  {Berkelbach}}, \bibinfo {author} {\bibfnamefont {H.}~\bibnamefont {Hill}},
  \bibinfo {author} {\bibfnamefont {A.}~\bibnamefont {Rigosi}}, \bibinfo
  {author} {\bibfnamefont {Y.}~\bibnamefont {Li}}, \bibinfo {author}
  {\bibfnamefont {O.}~\bibnamefont {Aslan}}, \bibinfo {author} {\bibfnamefont
  {D.}~\bibnamefont {Reichman}}, \bibinfo {author} {\bibfnamefont
  {M.}~\bibnamefont {Hybertsen}}, \ and\ \bibinfo {author} {\bibfnamefont
  {T.}~\bibnamefont {Heinz}},\ }\href@noop {} {\bibfield  {journal} {\bibinfo
  {journal} {Phys. Rev. Lett.}\ }\textbf {\bibinfo {volume} {113}},\ \bibinfo
  {pages} {076802} (\bibinfo {year} {2014})}\BibitemShut {NoStop}%
\bibitem [{\citenamefont {Hill}\ \emph {et~al.}(2015)\citenamefont {Hill},
  \citenamefont {Rigosi}, \citenamefont {Roquelet}, \citenamefont {Chernikov},
  \citenamefont {Berkelbach}, \citenamefont {Reichman}, \citenamefont
  {Hybertsen}, \citenamefont {Brus},\ and\ \citenamefont
  {Heinz}}]{RydbergStates_MoS2_Heinz_Nanolett2015}%
  \BibitemOpen
  \bibfield  {author} {\bibinfo {author} {\bibfnamefont {H.}~\bibnamefont
  {Hill}}, \bibinfo {author} {\bibfnamefont {A.}~\bibnamefont {Rigosi}},
  \bibinfo {author} {\bibfnamefont {C.}~\bibnamefont {Roquelet}}, \bibinfo
  {author} {\bibfnamefont {A.}~\bibnamefont {Chernikov}}, \bibinfo {author}
  {\bibfnamefont {T.}~\bibnamefont {Berkelbach}}, \bibinfo {author}
  {\bibfnamefont {D.}~\bibnamefont {Reichman}}, \bibinfo {author}
  {\bibfnamefont {M.}~\bibnamefont {Hybertsen}}, \bibinfo {author}
  {\bibfnamefont {L.}~\bibnamefont {Brus}}, \ and\ \bibinfo {author}
  {\bibfnamefont {T.}~\bibnamefont {Heinz}},\ }\href@noop {} {\bibfield
  {journal} {\bibinfo  {journal} {Nano Lett.}\ }\textbf {\bibinfo {volume}
  {15}},\ \bibinfo {pages} {2992} (\bibinfo {year} {2015})}\BibitemShut
  {NoStop}%
\bibitem [{\citenamefont {Schaibley}\ \emph {et~al.}(2016)\citenamefont
  {Schaibley}, \citenamefont {Yu}, \citenamefont {Clark}, \citenamefont
  {Rivera}, \citenamefont {Ross}, \citenamefont {Seyler}, \citenamefont {Yao},\
  and\ \citenamefont {Xu}}]{Valleytronics_XuReview_NatureReviews_2016}%
  \BibitemOpen
  \bibfield  {author} {\bibinfo {author} {\bibfnamefont {J.}~\bibnamefont
  {Schaibley}}, \bibinfo {author} {\bibfnamefont {H.}~\bibnamefont {Yu}},
  \bibinfo {author} {\bibfnamefont {G.}~\bibnamefont {Clark}}, \bibinfo
  {author} {\bibfnamefont {P.}~\bibnamefont {Rivera}}, \bibinfo {author}
  {\bibfnamefont {J.}~\bibnamefont {Ross}}, \bibinfo {author} {\bibfnamefont
  {K.}~\bibnamefont {Seyler}}, \bibinfo {author} {\bibfnamefont
  {W.}~\bibnamefont {Yao}}, \ and\ \bibinfo {author} {\bibfnamefont
  {X.}~\bibnamefont {Xu}},\ }\href@noop {} {\bibfield  {journal} {\bibinfo
  {journal} {Nat. Rev. Mater.}\ }\textbf {\bibinfo {volume} {1}} (\bibinfo
  {year} {2016})}\BibitemShut {NoStop}%
\bibitem [{\citenamefont {Yin}\ \emph {et~al.}(2012)\citenamefont {Yin},
  \citenamefont {Li}, \citenamefont {Li}, \citenamefont {Jiang}, \citenamefont
  {Shi}, \citenamefont {Sun}, \citenamefont {Lu}, \citenamefont {Zhang},
  \citenamefont {Chen},\ and\ \citenamefont
  {Zhang}}]{MoS2_phototransistor_ACSNano2012}%
  \BibitemOpen
  \bibfield  {author} {\bibinfo {author} {\bibfnamefont {Z.}~\bibnamefont
  {Yin}}, \bibinfo {author} {\bibfnamefont {H.}~\bibnamefont {Li}}, \bibinfo
  {author} {\bibfnamefont {H.}~\bibnamefont {Li}}, \bibinfo {author}
  {\bibfnamefont {L.}~\bibnamefont {Jiang}}, \bibinfo {author} {\bibfnamefont
  {Y.}~\bibnamefont {Shi}}, \bibinfo {author} {\bibfnamefont {Y.}~\bibnamefont
  {Sun}}, \bibinfo {author} {\bibfnamefont {G.}~\bibnamefont {Lu}}, \bibinfo
  {author} {\bibfnamefont {Q.}~\bibnamefont {Zhang}}, \bibinfo {author}
  {\bibfnamefont {X.}~\bibnamefont {Chen}}, \ and\ \bibinfo {author}
  {\bibfnamefont {H.}~\bibnamefont {Zhang}},\ }\href@noop {} {\bibfield
  {journal} {\bibinfo  {journal} {ACS Nano}\ }\textbf {\bibinfo {volume} {6}},\
  \bibinfo {pages} {74} (\bibinfo {year} {2012})}\BibitemShut {NoStop}%
\bibitem [{\citenamefont {Baugher}\ \emph {et~al.}(2014)\citenamefont
  {Baugher}, \citenamefont {Churchill}, \citenamefont {Yang},\ and\
  \citenamefont {Jarillo-Herrero}}]{WSe2OptoElec}%
  \BibitemOpen
  \bibfield  {author} {\bibinfo {author} {\bibfnamefont {B.~W.}\ \bibnamefont
  {Baugher}}, \bibinfo {author} {\bibfnamefont {H.~O.}\ \bibnamefont
  {Churchill}}, \bibinfo {author} {\bibfnamefont {Y.}~\bibnamefont {Yang}}, \
  and\ \bibinfo {author} {\bibfnamefont {P.}~\bibnamefont {Jarillo-Herrero}},\
  }\href@noop {} {\bibfield  {journal} {\bibinfo  {journal} {Nat.
  Nanotechnol.}\ }\textbf {\bibinfo {volume} {9}},\ \bibinfo {pages} {262}
  (\bibinfo {year} {2014})}\BibitemShut {NoStop}%
\bibitem [{\citenamefont {Ross}\ \emph {et~al.}(2014)\citenamefont {Ross},
  \citenamefont {Klement}, \citenamefont {Jones}, \citenamefont {Ghimire},
  \citenamefont {Yan}, \citenamefont {Mandrus}, \citenamefont {Taniguchi},
  \citenamefont {Watanabe}, \citenamefont {Kitamura}, \citenamefont {Yao},
  \citenamefont {Cobden},\ and\ \citenamefont {Xu}}]{WSe2LED}%
  \BibitemOpen
  \bibfield  {author} {\bibinfo {author} {\bibfnamefont {J.}~\bibnamefont
  {Ross}}, \bibinfo {author} {\bibfnamefont {P.}~\bibnamefont {Klement}},
  \bibinfo {author} {\bibfnamefont {A.}~\bibnamefont {Jones}}, \bibinfo
  {author} {\bibfnamefont {N.}~\bibnamefont {Ghimire}}, \bibinfo {author}
  {\bibfnamefont {J.}~\bibnamefont {Yan}}, \bibinfo {author} {\bibfnamefont
  {D.}~\bibnamefont {Mandrus}}, \bibinfo {author} {\bibfnamefont
  {T.}~\bibnamefont {Taniguchi}}, \bibinfo {author} {\bibfnamefont
  {K.}~\bibnamefont {Watanabe}}, \bibinfo {author} {\bibfnamefont
  {K.}~\bibnamefont {Kitamura}}, \bibinfo {author} {\bibfnamefont
  {W.}~\bibnamefont {Yao}}, \bibinfo {author} {\bibfnamefont {D.}~\bibnamefont
  {Cobden}}, \ and\ \bibinfo {author} {\bibfnamefont {X.}~\bibnamefont {Xu}},\
  }\href@noop {} {\bibfield  {journal} {\bibinfo  {journal} {Nat.
  Nanotechnol.}\ }\textbf {\bibinfo {volume} {9}},\ \bibinfo {pages} {268}
  (\bibinfo {year} {2014})}\BibitemShut {NoStop}%
\bibitem [{\citenamefont {Ye}\ \emph {et~al.}(2014)\citenamefont {Ye},
  \citenamefont {Ye}, \citenamefont {Gharghi}, \citenamefont {Zhu},
  \citenamefont {Zhao}, \citenamefont {Wang}, \citenamefont {Yin},\ and\
  \citenamefont {Zhang}}]{MoS2_ExcitonicLED_APL2014}%
  \BibitemOpen
  \bibfield  {author} {\bibinfo {author} {\bibfnamefont {Y.}~\bibnamefont
  {Ye}}, \bibinfo {author} {\bibfnamefont {Z.}~\bibnamefont {Ye}}, \bibinfo
  {author} {\bibfnamefont {M.}~\bibnamefont {Gharghi}}, \bibinfo {author}
  {\bibfnamefont {H.}~\bibnamefont {Zhu}}, \bibinfo {author} {\bibfnamefont
  {M.}~\bibnamefont {Zhao}}, \bibinfo {author} {\bibfnamefont {Y.}~\bibnamefont
  {Wang}}, \bibinfo {author} {\bibfnamefont {X.}~\bibnamefont {Yin}}, \ and\
  \bibinfo {author} {\bibfnamefont {X.}~\bibnamefont {Zhang}},\ }\href@noop {}
  {\bibfield  {journal} {\bibinfo  {journal} {Appl. Phys. Lett.}\ }\textbf
  {\bibinfo {volume} {104}} (\bibinfo {year} {2014})}\BibitemShut {NoStop}%
\bibitem [{\citenamefont {Sundaram}\ \emph {et~al.}(2013)\citenamefont
  {Sundaram}, \citenamefont {Engel}, \citenamefont {Lombardo}, \citenamefont
  {Krupke}, \citenamefont {Ferrari}, \citenamefont {Avouris},\ and\
  \citenamefont {Steiner}}]{MoS2_Electroluminescence_NanoLett2013}%
  \BibitemOpen
  \bibfield  {author} {\bibinfo {author} {\bibfnamefont {R.}~\bibnamefont
  {Sundaram}}, \bibinfo {author} {\bibfnamefont {M.}~\bibnamefont {Engel}},
  \bibinfo {author} {\bibfnamefont {A.}~\bibnamefont {Lombardo}}, \bibinfo
  {author} {\bibfnamefont {R.}~\bibnamefont {Krupke}}, \bibinfo {author}
  {\bibfnamefont {A.}~\bibnamefont {Ferrari}}, \bibinfo {author} {\bibfnamefont
  {P.}~\bibnamefont {Avouris}}, \ and\ \bibinfo {author} {\bibfnamefont
  {M.}~\bibnamefont {Steiner}},\ }\href@noop {} {\bibfield  {journal} {\bibinfo
   {journal} {Nano Lett.}\ }\textbf {\bibinfo {volume} {13}},\ \bibinfo {pages}
  {1416} (\bibinfo {year} {2013})}\BibitemShut {NoStop}%
\bibitem [{\citenamefont {Desai}\ \emph {et~al.}(2016)\citenamefont {Desai},
  \citenamefont {Madhvapathy}, \citenamefont {Sachid}, \citenamefont {Llinas},
  \citenamefont {Wang}, \citenamefont {Ahn}, \citenamefont {Pitner},
  \citenamefont {Kim}, \citenamefont {Bokor}, \citenamefont {Hu}, \citenamefont
  {Wong},\ and\ \citenamefont {Javey}}]{MoS2_transistor_1nmGate}%
  \BibitemOpen
  \bibfield  {author} {\bibinfo {author} {\bibfnamefont {S.}~\bibnamefont
  {Desai}}, \bibinfo {author} {\bibfnamefont {S.}~\bibnamefont {Madhvapathy}},
  \bibinfo {author} {\bibfnamefont {A.}~\bibnamefont {Sachid}}, \bibinfo
  {author} {\bibfnamefont {J.}~\bibnamefont {Llinas}}, \bibinfo {author}
  {\bibfnamefont {Q.}~\bibnamefont {Wang}}, \bibinfo {author} {\bibfnamefont
  {G.}~\bibnamefont {Ahn}}, \bibinfo {author} {\bibfnamefont {G.}~\bibnamefont
  {Pitner}}, \bibinfo {author} {\bibfnamefont {M.}~\bibnamefont {Kim}},
  \bibinfo {author} {\bibfnamefont {J.}~\bibnamefont {Bokor}}, \bibinfo
  {author} {\bibfnamefont {C.}~\bibnamefont {Hu}}, \bibinfo {author}
  {\bibfnamefont {H.-S.~P.}\ \bibnamefont {Wong}}, \ and\ \bibinfo {author}
  {\bibfnamefont {A.}~\bibnamefont {Javey}},\ }\href@noop {} {\bibfield
  {journal} {\bibinfo  {journal} {Science}\ }\textbf {\bibinfo {volume} {354}}
  (\bibinfo {year} {2016})}\BibitemShut {NoStop}%
\bibitem [{\citenamefont {Amani}\ \emph {et~al.}(2015)\citenamefont {Amani},
  \citenamefont {Lien}, \citenamefont {Kiriya}, \citenamefont {Xiao},
  \citenamefont {Azcatl}, \citenamefont {Noh}, \citenamefont {Madhvapathy},
  \citenamefont {Addou}, \citenamefont {KC}, \citenamefont {Dubey},
  \citenamefont {Cho}, \citenamefont {Wallace}, \citenamefont {Lee},
  \citenamefont {He}, \citenamefont {Ager~III}, \citenamefont {Zhang},
  \citenamefont {Yablonovitch},\ and\ \citenamefont
  {Javey}}]{NearUnityPL_QY_MoS2_Amani_etal_Science2015}%
  \BibitemOpen
  \bibfield  {author} {\bibinfo {author} {\bibfnamefont {M.}~\bibnamefont
  {Amani}}, \bibinfo {author} {\bibfnamefont {D.-H.}\ \bibnamefont {Lien}},
  \bibinfo {author} {\bibfnamefont {D.}~\bibnamefont {Kiriya}}, \bibinfo
  {author} {\bibfnamefont {J.}~\bibnamefont {Xiao}}, \bibinfo {author}
  {\bibfnamefont {A.}~\bibnamefont {Azcatl}}, \bibinfo {author} {\bibfnamefont
  {J.}~\bibnamefont {Noh}}, \bibinfo {author} {\bibfnamefont {S.}~\bibnamefont
  {Madhvapathy}}, \bibinfo {author} {\bibfnamefont {R.}~\bibnamefont {Addou}},
  \bibinfo {author} {\bibfnamefont {S.}~\bibnamefont {KC}}, \bibinfo {author}
  {\bibfnamefont {M.}~\bibnamefont {Dubey}}, \bibinfo {author} {\bibfnamefont
  {K.}~\bibnamefont {Cho}}, \bibinfo {author} {\bibfnamefont {R.}~\bibnamefont
  {Wallace}}, \bibinfo {author} {\bibfnamefont {S.-C.}\ \bibnamefont {Lee}},
  \bibinfo {author} {\bibfnamefont {J.-H.}\ \bibnamefont {He}}, \bibinfo
  {author} {\bibfnamefont {J.}~\bibnamefont {Ager~III}}, \bibinfo {author}
  {\bibfnamefont {X.}~\bibnamefont {Zhang}}, \bibinfo {author} {\bibfnamefont
  {E.}~\bibnamefont {Yablonovitch}}, \ and\ \bibinfo {author} {\bibfnamefont
  {A.}~\bibnamefont {Javey}},\ }\href@noop {} {\bibfield  {journal} {\bibinfo
  {journal} {Science}\ }\textbf {\bibinfo {volume} {350}} (\bibinfo {year}
  {2015})}\BibitemShut {NoStop}%
\bibitem [{\citenamefont {Dufferwiel}\ \emph {et~al.}(2015)\citenamefont
  {Dufferwiel}, \citenamefont {Schwarz}, \citenamefont {Withers}, \citenamefont
  {Trichet}, \citenamefont {Li}, \citenamefont {Sich}, \citenamefont {Del
  Pozo-Zamudio}, \citenamefont {Clark}, \citenamefont {Nalitov}, \citenamefont
  {Solnyshkov}, \citenamefont {Malpuech}, \citenamefont {Novoselov},
  \citenamefont {Smith}, \citenamefont {Skolnick}, \citenamefont
  {Krizhanovskii},\ and\ \citenamefont
  {Tartakovskii}}]{excitonPolaritons_heterostructures_microcav}%
  \BibitemOpen
  \bibfield  {author} {\bibinfo {author} {\bibfnamefont {S.}~\bibnamefont
  {Dufferwiel}}, \bibinfo {author} {\bibfnamefont {S.}~\bibnamefont {Schwarz}},
  \bibinfo {author} {\bibfnamefont {F.}~\bibnamefont {Withers}}, \bibinfo
  {author} {\bibfnamefont {A.}~\bibnamefont {Trichet}}, \bibinfo {author}
  {\bibfnamefont {F.}~\bibnamefont {Li}}, \bibinfo {author} {\bibfnamefont
  {M.}~\bibnamefont {Sich}}, \bibinfo {author} {\bibfnamefont {O.}~\bibnamefont
  {Del Pozo-Zamudio}}, \bibinfo {author} {\bibfnamefont {C.}~\bibnamefont
  {Clark}}, \bibinfo {author} {\bibfnamefont {A.}~\bibnamefont {Nalitov}},
  \bibinfo {author} {\bibfnamefont {D.}~\bibnamefont {Solnyshkov}}, \bibinfo
  {author} {\bibfnamefont {G.}~\bibnamefont {Malpuech}}, \bibinfo {author}
  {\bibfnamefont {K.}~\bibnamefont {Novoselov}}, \bibinfo {author}
  {\bibfnamefont {J.}~\bibnamefont {Smith}}, \bibinfo {author} {\bibfnamefont
  {M.}~\bibnamefont {Skolnick}}, \bibinfo {author} {\bibfnamefont
  {D.}~\bibnamefont {Krizhanovskii}}, \ and\ \bibinfo {author} {\bibfnamefont
  {A.}~\bibnamefont {Tartakovskii}},\ }\href@noop {} {\bibfield  {journal}
  {\bibinfo  {journal} {Nat. Commun.}\ }\textbf {\bibinfo {volume} {6}}
  (\bibinfo {year} {2015})}\BibitemShut {NoStop}%
\bibitem [{\citenamefont {Palummo}\ \emph {et~al.}(2015)\citenamefont
  {Palummo}, \citenamefont {Bernardi},\ and\ \citenamefont
  {Grossman}}]{TMDC_ExcitonLifetime_Grossman_2015Nanolett}%
  \BibitemOpen
  \bibfield  {author} {\bibinfo {author} {\bibfnamefont {M.}~\bibnamefont
  {Palummo}}, \bibinfo {author} {\bibfnamefont {M.}~\bibnamefont {Bernardi}}, \
  and\ \bibinfo {author} {\bibfnamefont {J.}~\bibnamefont {Grossman}},\
  }\href@noop {} {\bibfield  {journal} {\bibinfo  {journal} {Nano Lett.}\
  }\textbf {\bibinfo {volume} {15}},\ \bibinfo {pages} {2794} (\bibinfo {year}
  {2015})}\BibitemShut {NoStop}%
\bibitem [{\citenamefont {Moody}\ \emph {et~al.}(2015)\citenamefont {Moody},
  \citenamefont {Dass}, \citenamefont {Hao}, \citenamefont {Chen},
  \citenamefont {Li}, \citenamefont {Singh}, \citenamefont {Tran},
  \citenamefont {Clark}, \citenamefont {Xu}, \citenamefont {Berghauser},
  \citenamefont {Malic}, \citenamefont {Knorr},\ and\ \citenamefont
  {Li}}]{MoS2_2DElectronicSpectroscopy}%
  \BibitemOpen
  \bibfield  {author} {\bibinfo {author} {\bibfnamefont {G.}~\bibnamefont
  {Moody}}, \bibinfo {author} {\bibfnamefont {C.}~\bibnamefont {Dass}},
  \bibinfo {author} {\bibfnamefont {K.}~\bibnamefont {Hao}}, \bibinfo {author}
  {\bibfnamefont {C.-H.}\ \bibnamefont {Chen}}, \bibinfo {author}
  {\bibfnamefont {L.-J.}\ \bibnamefont {Li}}, \bibinfo {author} {\bibfnamefont
  {A.}~\bibnamefont {Singh}}, \bibinfo {author} {\bibfnamefont
  {K.}~\bibnamefont {Tran}}, \bibinfo {author} {\bibfnamefont {G.}~\bibnamefont
  {Clark}}, \bibinfo {author} {\bibfnamefont {X.}~\bibnamefont {Xu}}, \bibinfo
  {author} {\bibfnamefont {G.}~\bibnamefont {Berghauser}}, \bibinfo {author}
  {\bibfnamefont {E.}~\bibnamefont {Malic}}, \bibinfo {author} {\bibfnamefont
  {A.}~\bibnamefont {Knorr}}, \ and\ \bibinfo {author} {\bibfnamefont
  {X.}~\bibnamefont {Li}},\ }\href@noop {} {\bibfield  {journal} {\bibinfo
  {journal} {Nat. Commun.}\ }\textbf {\bibinfo {volume} {6}} (\bibinfo {year}
  {2015})}\BibitemShut {NoStop}%
\bibitem [{\citenamefont {Korn}\ \emph {et~al.}(2011)\citenamefont {Korn},
  \citenamefont {Heydrich}, \citenamefont {Himer}, \citenamefont {Schmutzler},\
  and\ \citenamefont {Schuller}}]{UltrfastMoS2_TRPL_LowTemp_APL2011}%
  \BibitemOpen
  \bibfield  {author} {\bibinfo {author} {\bibfnamefont {T.}~\bibnamefont
  {Korn}}, \bibinfo {author} {\bibfnamefont {S.}~\bibnamefont {Heydrich}},
  \bibinfo {author} {\bibfnamefont {M.}~\bibnamefont {Himer}}, \bibinfo
  {author} {\bibfnamefont {J.}~\bibnamefont {Schmutzler}}, \ and\ \bibinfo
  {author} {\bibfnamefont {C.}~\bibnamefont {Schuller}},\ }\href@noop {}
  {\bibfield  {journal} {\bibinfo  {journal} {Appl. Phys. Lett.}\ }\textbf
  {\bibinfo {volume} {99}} (\bibinfo {year} {2011})}\BibitemShut {NoStop}%
\bibitem [{\citenamefont {Wang}\ \emph {et~al.}(2015)\citenamefont {Wang},
  \citenamefont {Strait}, \citenamefont {Zhang}, \citenamefont {Chan},
  \citenamefont {Manolatou}, \citenamefont {Tiwari},\ and\ \citenamefont
  {Rana}}]{FastAnnihilationDefects_TMDS_PRB2015}%
  \BibitemOpen
  \bibfield  {author} {\bibinfo {author} {\bibfnamefont {H.}~\bibnamefont
  {Wang}}, \bibinfo {author} {\bibfnamefont {J.}~\bibnamefont {Strait}},
  \bibinfo {author} {\bibfnamefont {C.}~\bibnamefont {Zhang}}, \bibinfo
  {author} {\bibfnamefont {W.}~\bibnamefont {Chan}}, \bibinfo {author}
  {\bibfnamefont {C.}~\bibnamefont {Manolatou}}, \bibinfo {author}
  {\bibfnamefont {S.}~\bibnamefont {Tiwari}}, \ and\ \bibinfo {author}
  {\bibfnamefont {F.}~\bibnamefont {Rana}},\ }\href@noop {} {\bibfield
  {journal} {\bibinfo  {journal} {Phys. Rev. B}\ }\textbf {\bibinfo {volume}
  {91}} (\bibinfo {year} {2015})}\BibitemShut {NoStop}%
\bibitem [{\citenamefont {Sun}\ \emph {et~al.}(2014)\citenamefont {Sun},
  \citenamefont {Rao}, \citenamefont {Reider}, \citenamefont {Chen},
  \citenamefont {You}, \citenamefont {Brezin}, \citenamefont {Harutyunyan},\
  and\ \citenamefont {Heinz}}]{UltrafastMoS2_XXAnnihilation_NL2014}%
  \BibitemOpen
  \bibfield  {author} {\bibinfo {author} {\bibfnamefont {D.}~\bibnamefont
  {Sun}}, \bibinfo {author} {\bibfnamefont {Y.}~\bibnamefont {Rao}}, \bibinfo
  {author} {\bibfnamefont {G.}~\bibnamefont {Reider}}, \bibinfo {author}
  {\bibfnamefont {G.}~\bibnamefont {Chen}}, \bibinfo {author} {\bibfnamefont
  {Y.}~\bibnamefont {You}}, \bibinfo {author} {\bibfnamefont {L.}~\bibnamefont
  {Brezin}}, \bibinfo {author} {\bibfnamefont {A.}~\bibnamefont {Harutyunyan}},
  \ and\ \bibinfo {author} {\bibfnamefont {T.}~\bibnamefont {Heinz}},\
  }\href@noop {} {\bibfield  {journal} {\bibinfo  {journal} {Nano Lett.}\
  }\textbf {\bibinfo {volume} {14}} (\bibinfo {year} {2014})}\BibitemShut
  {NoStop}%
\bibitem [{\citenamefont {Zhang}\ \emph {et~al.}(2015)\citenamefont {Zhang},
  \citenamefont {You}, \citenamefont {Zhao},\ and\ \citenamefont
  {Heinz}}]{DarkExcitonsWSe2_HeinzPRL2015}%
  \BibitemOpen
  \bibfield  {author} {\bibinfo {author} {\bibfnamefont {X.-X.}\ \bibnamefont
  {Zhang}}, \bibinfo {author} {\bibfnamefont {Y.}~\bibnamefont {You}}, \bibinfo
  {author} {\bibfnamefont {S.}~\bibnamefont {Zhao}}, \ and\ \bibinfo {author}
  {\bibfnamefont {T.}~\bibnamefont {Heinz}},\ }\href@noop {} {\bibfield
  {journal} {\bibinfo  {journal} {Phys. Rev. Lett.}\ }\textbf {\bibinfo
  {volume} {115}} (\bibinfo {year} {2015})}\BibitemShut {NoStop}%
\bibitem [{\citenamefont
  {Andreani}(1991)}]{IntrinsicRadDecayTimes_quantumWells}%
  \BibitemOpen
  \bibfield  {author} {\bibinfo {author} {\bibfnamefont {L.}~\bibnamefont
  {Andreani}},\ }\href@noop {} {\bibfield  {journal} {\bibinfo  {journal}
  {Solid State Commun.}\ }\textbf {\bibinfo {volume} {9}} (\bibinfo {year}
  {1991})}\BibitemShut {NoStop}%
\bibitem [{\citenamefont {Glazov}\ \emph {et~al.}(2015)\citenamefont {Glazov},
  \citenamefont {Ivchenko}, \citenamefont {Wang}, \citenamefont {Amand},
  \citenamefont {Marie}, \citenamefont {Urbaszek},\ and\ \citenamefont
  {Liu}}]{IntrinsicRadDecayTimes_2DMaterials}%
  \BibitemOpen
  \bibfield  {author} {\bibinfo {author} {\bibfnamefont {M.}~\bibnamefont
  {Glazov}}, \bibinfo {author} {\bibfnamefont {E.}~\bibnamefont {Ivchenko}},
  \bibinfo {author} {\bibfnamefont {G.}~\bibnamefont {Wang}}, \bibinfo {author}
  {\bibfnamefont {T.}~\bibnamefont {Amand}}, \bibinfo {author} {\bibfnamefont
  {X.}~\bibnamefont {Marie}}, \bibinfo {author} {\bibfnamefont
  {B.}~\bibnamefont {Urbaszek}}, \ and\ \bibinfo {author} {\bibfnamefont
  {B.}~\bibnamefont {Liu}},\ }\href@noop {} {\bibfield  {journal} {\bibinfo
  {journal} {Phys. Status Solidi. B}\ }\textbf {\bibinfo {volume} {252}}
  (\bibinfo {year} {2015})}\BibitemShut {NoStop}%
\bibitem [{\citenamefont {Robert}\ \emph {et~al.}(2016)\citenamefont {Robert},
  \citenamefont {Lagarde}, \citenamefont {Cadiz}, \citenamefont {Wang},
  \citenamefont {Lassagne}, \citenamefont {Amand}, \citenamefont {Balocchi},
  \citenamefont {Renucci}, \citenamefont {Tongay}, \citenamefont {Urbaszek},\
  and\ \citenamefont {Marie}}]{2D_Mat_RadLifetimes_PRB}%
  \BibitemOpen
  \bibfield  {author} {\bibinfo {author} {\bibfnamefont {C.}~\bibnamefont
  {Robert}}, \bibinfo {author} {\bibfnamefont {D.}~\bibnamefont {Lagarde}},
  \bibinfo {author} {\bibfnamefont {F.}~\bibnamefont {Cadiz}}, \bibinfo
  {author} {\bibfnamefont {G.}~\bibnamefont {Wang}}, \bibinfo {author}
  {\bibfnamefont {B.}~\bibnamefont {Lassagne}}, \bibinfo {author}
  {\bibfnamefont {T.}~\bibnamefont {Amand}}, \bibinfo {author} {\bibfnamefont
  {A.}~\bibnamefont {Balocchi}}, \bibinfo {author} {\bibfnamefont
  {P.}~\bibnamefont {Renucci}}, \bibinfo {author} {\bibfnamefont
  {S.}~\bibnamefont {Tongay}}, \bibinfo {author} {\bibfnamefont
  {B.}~\bibnamefont {Urbaszek}}, \ and\ \bibinfo {author} {\bibfnamefont
  {X.}~\bibnamefont {Marie}},\ }\href@noop {} {\bibfield  {journal} {\bibinfo
  {journal} {Phys. Rev. B}\ }\textbf {\bibinfo {volume} {93}} (\bibinfo {year}
  {2016})}\BibitemShut {NoStop}%
\bibitem [{\citenamefont {Li}\ \emph {et~al.}(2014)\citenamefont {Li},
  \citenamefont {Chernikov}, \citenamefont {Zhang}, \citenamefont {Rigosi},
  \citenamefont {Hill}, \citenamefont {Van~der Zande}, \citenamefont {Chenet},
  \citenamefont {Shih}, \citenamefont {Hone},\ and\ \citenamefont
  {Heinz}}]{TMD_dielectric_measured_HeinzPRB2014}%
  \BibitemOpen
  \bibfield  {author} {\bibinfo {author} {\bibfnamefont {Y.}~\bibnamefont
  {Li}}, \bibinfo {author} {\bibfnamefont {A.}~\bibnamefont {Chernikov}},
  \bibinfo {author} {\bibfnamefont {X.}~\bibnamefont {Zhang}}, \bibinfo
  {author} {\bibfnamefont {A.}~\bibnamefont {Rigosi}}, \bibinfo {author}
  {\bibfnamefont {H.}~\bibnamefont {Hill}}, \bibinfo {author} {\bibfnamefont
  {A.}~\bibnamefont {Van~der Zande}}, \bibinfo {author} {\bibfnamefont
  {D.}~\bibnamefont {Chenet}}, \bibinfo {author} {\bibfnamefont {E.-M.}\
  \bibnamefont {Shih}}, \bibinfo {author} {\bibfnamefont {J.}~\bibnamefont
  {Hone}}, \ and\ \bibinfo {author} {\bibfnamefont {T.}~\bibnamefont {Heinz}},\
  }\href@noop {} {\bibfield  {journal} {\bibinfo  {journal} {Phys. Rev. B}\
  }\textbf {\bibinfo {volume} {90}} (\bibinfo {year} {2014})}\BibitemShut
  {NoStop}%
\bibitem [{\citenamefont {Mayers}\ \emph {et~al.}(2015)\citenamefont {Mayers},
  \citenamefont {Berkelbach}, \citenamefont {Hybertsen},\ and\ \citenamefont
  {Reichman}}]{CalcBindingEnergy_X_XX_MoS2_PRB2015}%
  \BibitemOpen
  \bibfield  {author} {\bibinfo {author} {\bibfnamefont {Z.}~\bibnamefont
  {Mayers}}, \bibinfo {author} {\bibfnamefont {T.}~\bibnamefont {Berkelbach}},
  \bibinfo {author} {\bibfnamefont {M.}~\bibnamefont {Hybertsen}}, \ and\
  \bibinfo {author} {\bibfnamefont {D.}~\bibnamefont {Reichman}},\ }\href@noop
  {} {\bibfield  {journal} {\bibinfo  {journal} {Phys. Rev. B}\ }\textbf
  {\bibinfo {volume} {92}},\ \bibinfo {pages} {161404} (\bibinfo {year}
  {2015})}\BibitemShut {NoStop}%
\bibitem [{\citenamefont {Zhou}\ \emph {et~al.}(2013)\citenamefont {Zhou},
  \citenamefont {Zou}, \citenamefont {Najmaei}, \citenamefont {Liu},
  \citenamefont {Shi}, \citenamefont {Kong}, \citenamefont {Lou}, \citenamefont
  {Ajayan}, \citenamefont {Yakobson},\ and\ \citenamefont
  {Idrobo}}]{IntrinsicStructural_MonolayerMoS2_NanoLett}%
  \BibitemOpen
  \bibfield  {author} {\bibinfo {author} {\bibfnamefont {W.}~\bibnamefont
  {Zhou}}, \bibinfo {author} {\bibfnamefont {X.}~\bibnamefont {Zou}}, \bibinfo
  {author} {\bibfnamefont {S.}~\bibnamefont {Najmaei}}, \bibinfo {author}
  {\bibfnamefont {Z.}~\bibnamefont {Liu}}, \bibinfo {author} {\bibfnamefont
  {Y.}~\bibnamefont {Shi}}, \bibinfo {author} {\bibfnamefont {J.}~\bibnamefont
  {Kong}}, \bibinfo {author} {\bibfnamefont {J.}~\bibnamefont {Lou}}, \bibinfo
  {author} {\bibfnamefont {P.}~\bibnamefont {Ajayan}}, \bibinfo {author}
  {\bibfnamefont {B.}~\bibnamefont {Yakobson}}, \ and\ \bibinfo {author}
  {\bibfnamefont {J.-C.}\ \bibnamefont {Idrobo}},\ }\href@noop {} {\bibfield
  {journal} {\bibinfo  {journal} {Nano Lett.}\ }\textbf {\bibinfo {volume}
  {13}},\ \bibinfo {pages} {2615} (\bibinfo {year} {2013})}\BibitemShut
  {NoStop}%
\bibitem [{\citenamefont {Addou}\ \emph {et~al.}(2015)\citenamefont {Addou},
  \citenamefont {Colombo},\ and\ \citenamefont
  {Wallace}}]{MoS2_SurfaceDefectsOnNatural_ACSAMI_2015}%
  \BibitemOpen
  \bibfield  {author} {\bibinfo {author} {\bibfnamefont {R.}~\bibnamefont
  {Addou}}, \bibinfo {author} {\bibfnamefont {L.}~\bibnamefont {Colombo}}, \
  and\ \bibinfo {author} {\bibfnamefont {R.}~\bibnamefont {Wallace}},\
  }\href@noop {} {\bibfield  {journal} {\bibinfo  {journal} {ACS Appl. Mater.
  Interfaces}\ }\textbf {\bibinfo {volume} {7}},\ \bibinfo {pages} {11921}
  (\bibinfo {year} {2015})}\BibitemShut {NoStop}%
\bibitem [{\citenamefont {Amani}\ \emph {et~al.}(2016)\citenamefont {Amani},
  \citenamefont {Taheri}, \citenamefont {Addou}, \citenamefont {Ahn},
  \citenamefont {Kiriya}, \citenamefont {Lien}, \citenamefont {Ager~III},
  \citenamefont {Wallace},\ and\ \citenamefont
  {Javey}}]{Amani_TFSI_Nanoletters2016_followup}%
  \BibitemOpen
  \bibfield  {author} {\bibinfo {author} {\bibfnamefont {M.}~\bibnamefont
  {Amani}}, \bibinfo {author} {\bibfnamefont {P.}~\bibnamefont {Taheri}},
  \bibinfo {author} {\bibfnamefont {R.}~\bibnamefont {Addou}}, \bibinfo
  {author} {\bibfnamefont {G.}~\bibnamefont {Ahn}}, \bibinfo {author}
  {\bibfnamefont {D.}~\bibnamefont {Kiriya}}, \bibinfo {author} {\bibfnamefont
  {D.-H.}\ \bibnamefont {Lien}}, \bibinfo {author} {\bibfnamefont
  {J.}~\bibnamefont {Ager~III}}, \bibinfo {author} {\bibfnamefont
  {R.}~\bibnamefont {Wallace}}, \ and\ \bibinfo {author} {\bibfnamefont
  {A.}~\bibnamefont {Javey}},\ }\href@noop {} {\bibfield  {journal} {\bibinfo
  {journal} {Nano Lett.}\ }\textbf {\bibinfo {volume} {16}},\ \bibinfo {pages}
  {2786} (\bibinfo {year} {2016})}\BibitemShut {NoStop}%
\bibitem [{\citenamefont {Atallah}\ \emph {et~al.}(2017)\citenamefont
  {Atallah}, \citenamefont {Wang}, \citenamefont {Bosch}, \citenamefont {Seo},
  \citenamefont {Burke}, \citenamefont {Moneer}, \citenamefont {Zhu},
  \citenamefont {Theibault}, \citenamefont {Brus}, \citenamefont {Hone},\ and\
  \citenamefont {Zhu}}]{2D_ionicLiquid_XYZ_2017Nanolett}%
  \BibitemOpen
  \bibfield  {author} {\bibinfo {author} {\bibfnamefont {T.}~\bibnamefont
  {Atallah}}, \bibinfo {author} {\bibfnamefont {J.}~\bibnamefont {Wang}},
  \bibinfo {author} {\bibfnamefont {M.}~\bibnamefont {Bosch}}, \bibinfo
  {author} {\bibfnamefont {D.}~\bibnamefont {Seo}}, \bibinfo {author}
  {\bibfnamefont {R.}~\bibnamefont {Burke}}, \bibinfo {author} {\bibfnamefont
  {O.}~\bibnamefont {Moneer}}, \bibinfo {author} {\bibfnamefont
  {J.}~\bibnamefont {Zhu}}, \bibinfo {author} {\bibfnamefont {M.}~\bibnamefont
  {Theibault}}, \bibinfo {author} {\bibfnamefont {L.}~\bibnamefont {Brus}},
  \bibinfo {author} {\bibfnamefont {J.}~\bibnamefont {Hone}}, \ and\ \bibinfo
  {author} {\bibfnamefont {X.-Y.}\ \bibnamefont {Zhu}},\ }\href@noop {}
  {\bibfield  {journal} {\bibinfo  {journal} {J. Phys. Chem. Lett.}\ }\textbf
  {\bibinfo {volume} {8}} (\bibinfo {year} {2017})}\BibitemShut {NoStop}%
\bibitem [{\citenamefont {Kim}\ \emph {et~al.}(2017)\citenamefont {Kim},
  \citenamefont {Lien}, \citenamefont {Amani}, \citenamefont {Ager},\ and\
  \citenamefont {Javey}}]{CYTOPStability}%
  \BibitemOpen
  \bibfield  {author} {\bibinfo {author} {\bibfnamefont {H.}~\bibnamefont
  {Kim}}, \bibinfo {author} {\bibfnamefont {D.-H.}\ \bibnamefont {Lien}},
  \bibinfo {author} {\bibfnamefont {M.}~\bibnamefont {Amani}}, \bibinfo
  {author} {\bibfnamefont {J.}~\bibnamefont {Ager}}, \ and\ \bibinfo {author}
  {\bibfnamefont {A.}~\bibnamefont {Javey}},\ }\href@noop {} {\bibfield
  {journal} {\bibinfo  {journal} {ACS Nano}\ } (\bibinfo {year}
  {2017})}\BibitemShut {NoStop}%
\bibitem [{\citenamefont {Lee}\ \emph {et~al.}(2010)\citenamefont {Lee},
  \citenamefont {Yan}, \citenamefont {Brus}, \citenamefont {Heinz},
  \citenamefont {Hone},\ and\ \citenamefont {Ryu}}]{MoS2Raman}%
  \BibitemOpen
  \bibfield  {author} {\bibinfo {author} {\bibfnamefont {C.}~\bibnamefont
  {Lee}}, \bibinfo {author} {\bibfnamefont {H.}~\bibnamefont {Yan}}, \bibinfo
  {author} {\bibfnamefont {L.}~\bibnamefont {Brus}}, \bibinfo {author}
  {\bibfnamefont {T.}~\bibnamefont {Heinz}}, \bibinfo {author} {\bibfnamefont
  {J.}~\bibnamefont {Hone}}, \ and\ \bibinfo {author} {\bibfnamefont
  {S.}~\bibnamefont {Ryu}},\ }\href@noop {} {\bibfield  {journal} {\bibinfo
  {journal} {ACS Nano}\ }\textbf {\bibinfo {volume} {4}},\ \bibinfo {pages}
  {1695} (\bibinfo {year} {2010})}\BibitemShut {NoStop}%
\bibitem [{\citenamefont {Mooney}\ and\ \citenamefont
  {Kambhampati}(2013)}]{Jacobian_Transform}%
  \BibitemOpen
  \bibfield  {author} {\bibinfo {author} {\bibfnamefont {J.}~\bibnamefont
  {Mooney}}\ and\ \bibinfo {author} {\bibfnamefont {P.}~\bibnamefont
  {Kambhampati}},\ }\href@noop {} {\bibfield  {journal} {\bibinfo  {journal}
  {J. Phys. Chem. Lett.}\ }\textbf {\bibinfo {volume} {4}} (\bibinfo {year}
  {2013})}\BibitemShut {NoStop}%
\end{thebibliography}%

\FloatBarrier
\section{Supplementary Information}
\beginsupplement
\section{Trap States in  CVD-Grown MoS$_2$}
\FloatBarrier
\begin{figure}[h!]
\includegraphics[width=.5\columnwidth]{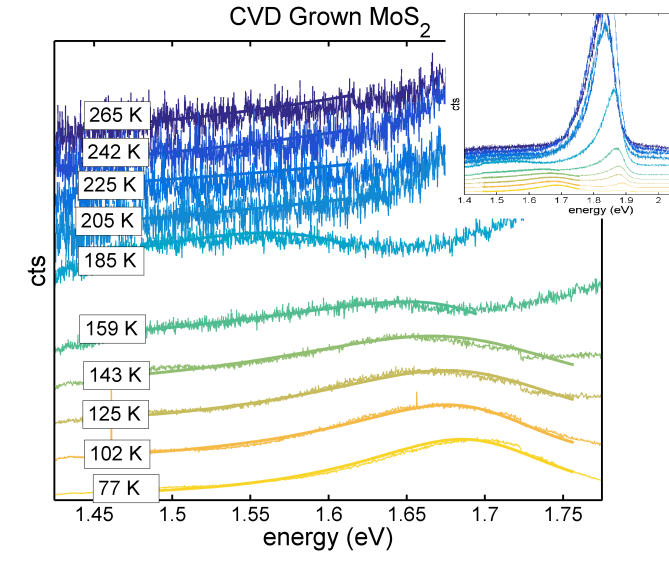}
\caption{Temperature dependent PL spectra from acid-treated CVD-grown MoS$_2$. The CVD-grown sample also exhibits trapped exciton emission at low temperature. The trapped exciton emission line shape differs reflecting a different energetic distribution of trap states.}
\label{FigS1}
\end{figure}

Multiple exfoliated samples as well as CVD grown samples exhibit the trapped exciton emission at low temperature. Notably, the trapped exciton emission varies in prominence from sample to sample, and the exact lineshape also varies. This suggests that the average area-density and energetic distribution of traps may vary sample to sample, supporting our hypothesis that the trap states are determined by the quality of the initial starting material rather than caused by the superacid treatment itself. Importantly, the dark traps influence exciton behavior even when not apparent in room temperature absorption and emission spectra. 

Here, spectra from a CVD grown sample are plotted in Fig~\ref{FigS1}. The trapped-exciton emission was fitted to an exponentially decaying density of trap states filled with a Fermi-Dirac distribution as in the main text, but the fitted parameter describing the energetic distribution of trap states was found to be $\alpha\sim 10~\mathrm{eV}^{-1}$ as opposed to the $\alpha\sim 5~\mathrm{eV}^{-1}$ found in the main text. This parameter, $\alpha$, parameterizes the density of states according to
\begin{align}
\rho \propto \exp \left[-\alpha\left(E-E_\mathrm{band~edge}\right)\right].
\end{align}

\FloatBarrier
\section{Numerical Modelling Details and Results}
\subsection{Steady State Density Dependence}
\FloatBarrier
\begin{figure}[h!]
\includegraphics[width=.6\columnwidth]{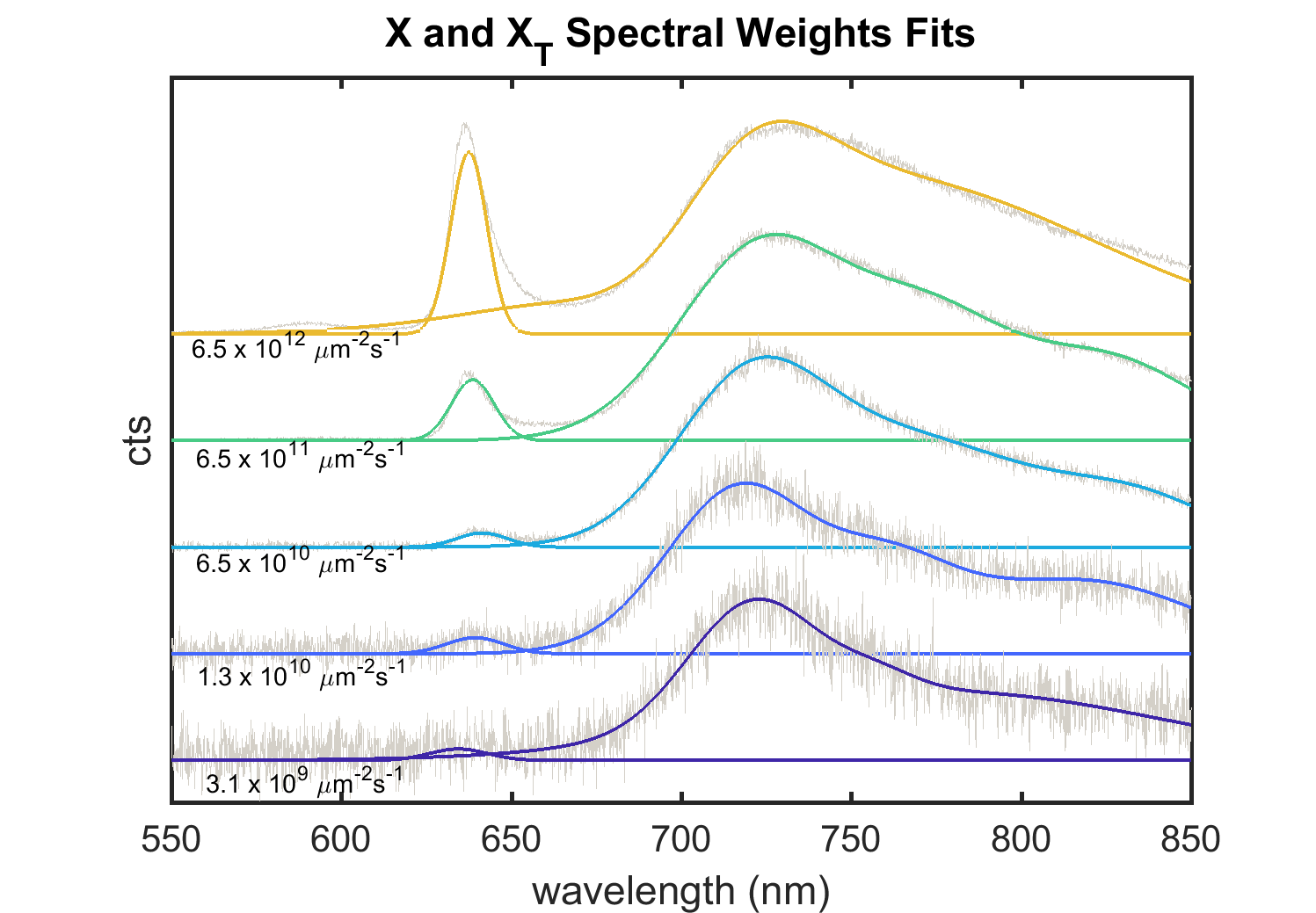}
\caption{Power-dependent PL at 77~K. Power-dependent PL spectra (light gray) were fitted to band edge and trapped exciton contributions (colored lines) to extract density-dependent steady state populations.}
\label{FigS2}
\end{figure}

Power-dependent spectra were collected at 77~K to extract the trapping rate, $k_\mathrm{trap}$, and the trap state density, $N_0$. The power-dependent spectra were analyzed to extract density-dependent steady-state populations of the band edge $[X]$ and trapped $[X_T]$ populations. The spectra and their fits are plotted in Fig.~\ref{FigS2}. After integrating the counts from each state and considering the two states’ markedly different radiative rates, the fraction of the total population at the band edge, $[X]/([X]+[X_T])$ was found and used for modelling.

The expected power-dependent equilibrium populations were predicted by considering Eqns. (3-4) from the main text reproduced here:
\begin{align}
\frac{\mathrm{d}[X]}{\mathrm{d}t}&= R_\mathrm{gen}-k_{\mathrm{rad}}^{\mathrm{band\mhyphen edge}}[X]
-k_{\mathrm{trap}}[X]\left(1-\frac{[X_T]}{N_0}\right)+k_{\mathrm{detrap}}[X_T]
\label{X_pop}\\
\frac{\mathrm{d}[X_T]}{\mathrm{d}t}&=-k_{\mathrm{rad}}^{\mathrm{trapped}}[X_T]
-k_{\mathrm{detrap}}[X_T]+k_{\mathrm{trap}}[X]\left(1-\frac{[X_T]}{N_0}\right).
\label{XT_pop}
\end{align}
$[X]$ and $[X_T]$ are the band edge and trapped exciton populations respectively. $R_\mathrm{gen}$ is the generation rate determined by the laser power, and $k_{\mathrm{rad}}^{\mathrm{band\mhyphen edge}}$, $k_{\mathrm{trap}}$, $k_{\mathrm{detrap}}$, $k_{\mathrm{rad}}^{\mathrm{trapped}}$ are the band-edge radiative rate, trapping rate, detrapping rate, and trapped exciton radiative rate respectively. In the CW PL spectroscopy experiment, the populations reach a steady-state, and their rates of change (the LHS of each equation) are zero. This establishes a set of two coupled equations in $[X]^\mathrm{eq}$ and $[X_T]^\mathrm{eq}$ with $R_{gen}$ determined by the laser power. The equations were solved numerically while varying $k_\mathrm{trap}$ and $N_0$ to reproduce the experimentally found fluence-dependent band-edge population fraction, $[X]/([X]+[X_T])$.

\FloatBarrier
\subsection{Transient ODE Modelling}
\FloatBarrier

\begin{figure}[h!]
\includegraphics[width=.7\columnwidth]{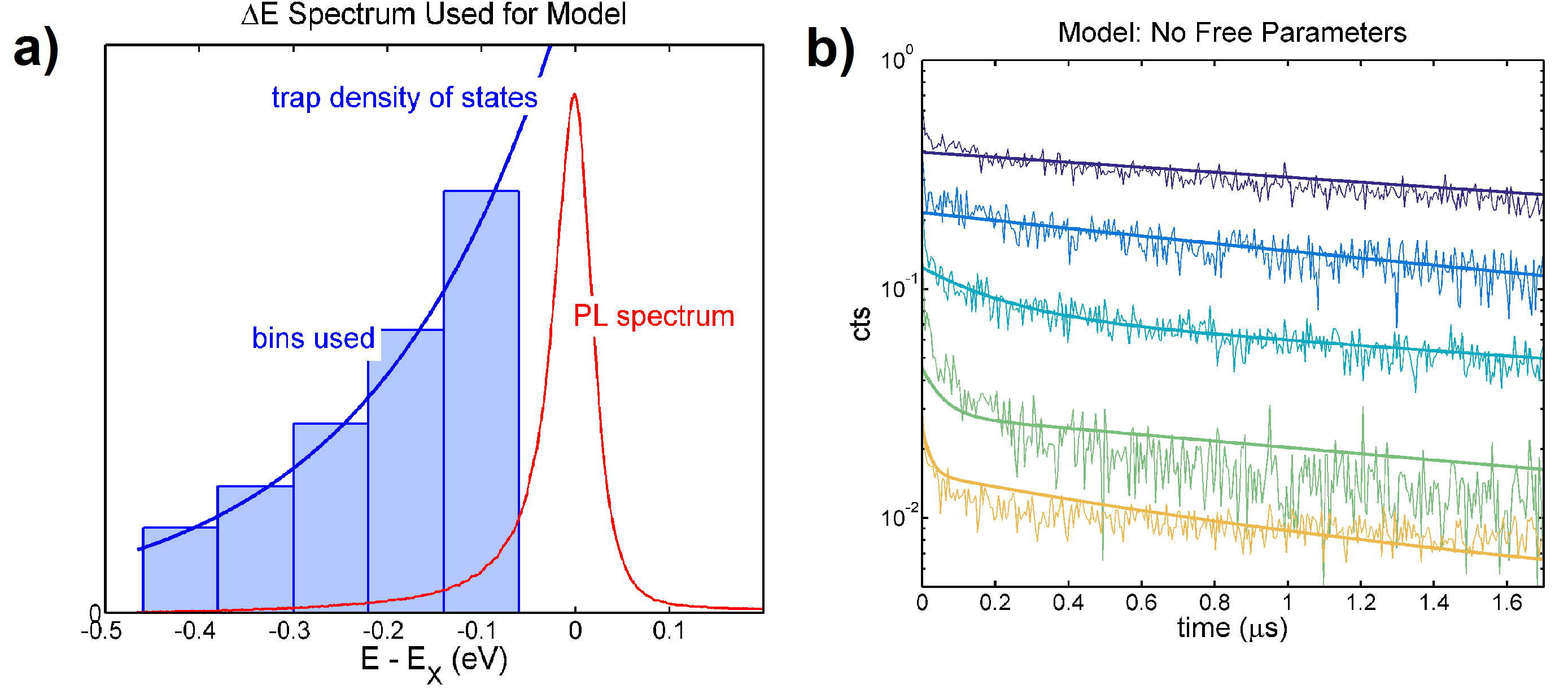}
\caption{Transient photoluminescence modelling. a) The trap state energy spectrum used in the numerical model is illustrated here. The PL spectrum is plotted in red with the x-axis energy centered about the band edge exciton energy, $E_X$. The density of trap states is exponentially decaying from the band edge with a slope $\alpha=5~\mathrm{eV}^{-1}$ found from fitting the PL spectra. This exponential decay is plotted in blue. The transient ODE model was solved five times at each temperature with trap state energies represented by the blue rectangles. The weight of each solution parameterized by a specific trap state energy was determined by the density of states. The five solutions for each temperature were added together to generate the model’s output. b) The experimental trapped exciton emission at multiple temperatures is plotted along with the model output, which was generated with no free parameters.}
\label{FigS3}
\end{figure}

The temperature-dependent trapped exciton emission (Fig.~3b in the main text and Fig.~\ref{FigS3}b here) was numerically modelled using Eqns. (\ref{X_pop} - \ref{XT_pop}). The time-dependent model was seeded with initial conditions $[X](t=0)=0$ and a trap state population determined by the experimental pulse energy. The pulse generated 100,000 total trapped excitations that were distributed over five energetic bins below the band gap. The excitations were partitioned into the bins according to the experimentally determined trap density of states. This partitioning is illustrated in Fig.~\ref{FigS3}a. Each set of initial conditions was numerically integrated and the solutions for each trap state energy were added together to generate a solution at each temperature to compare to the experimental data. With no free parameters, the model produces solutions that closely match the multi-exponential behavior shown in the data (Fig.~\ref{FigS3}b). The model captures behavior at low temperatures: trapped excitons don’t have sufficient thermal energy to escape to the band edge. Conversely, at higher temperatures, excitons trapped at different energies can be thermally promoted to the band edge with different average rates determined by the depth of their traps.
\FloatBarrier

\end{document}